\documentclass[
prc,%
10pt,%
final,%
notitlepage,%
oneside,%
twocolumn,%
nobibnotes,%
nofootinbib,
superscriptaddress,%
floatfix,%
floatfix,%
showkeys,%
showpacs]%
{revtex4}
\usepackage{color}
\usepackage{amsfonts}
\usepackage{amsbsy}
\usepackage{mathrsfs}
\usepackage{graphicx}
\def\lsim{\mathrel{\rlap{
\lower4pt\hbox{\hskip-3pt$\sim$}}
    \raise1pt\hbox{$<$}}}     
\def\gsim{\mathrel{\rlap{
\lower4pt\hbox{\hskip-3pt$\sim$}}
    \raise1pt\hbox{$>$}}}     

\begin{document}
\title{	
On Ambiguity of Definition of Shear and Spin-Hall Contributions to $\Lambda$ Polarization in Heavy-Ion Collisions
} 
\author{Yu. B. Ivanov}\thanks{e-mail: yivanov@theor.jinr.ru}
\affiliation{Bogoliubov Laboratory for Theoretical Physics, 
Joint Institute for Nuclear Research, Dubna 141980, Russia}
\affiliation{National Research Nuclear University "MEPhI", 
Moscow 115409, Russia}
\affiliation{National Research Centre "Kurchatov Institute",  Moscow 123182, Russia} 
\author{A. A. Soldatov}
\affiliation{National Research Nuclear University "MEPhI",
Moscow 115409, Russia}
\begin{abstract}
Recently proposed thermal-shear and spin-Hall contributions to the particle polarization in heavy-ion collisions 
are discussed. Alternative definitions of the thermal-shear contribution, i.e. those of 
Becattini-Buzzegoli-Palermo on the one hand and Liu-Yin on the other, 
are very similar in the midrapidity region while quite different at forward-backward rapidities, 
which are measured in fixed-target experiments. It is shown that the thermal-shear contribution to the global polarization 
with momentum averaging extended to all momenta is very different within these alternative definitions. 
The spin-Hall contribution to the global polarization, defined similarly to 
the Liu-Yin shear one, is identically zero, if averaging runs  
over all momenta. Only application of restrictive momentum acceptance 
and the boost (to $\Lambda$ rest frame) correction results in nonzero 
global spin-Hall polarization. 
If the spin-Hall contribution were defined similarly to Becattini-Buzzegoli-Palermo shear one, 
the global spin-Hall polarization would be non-zero even without any acceptance and the boost correction.
\end{abstract}
\maketitle

Non-central heavy-ion collisions at high energies are
characterized by a huge global angular momentum of the order of 10$^3$-10$^5 \hbar$, 
depending on the collision energy and centrality, e.g., see Refs. \cite{Ivanov:2019ern,Ivanov:2019wzg}.
Although a large part of the angular momentum is carried away
by the spectator nucleons, its sizable fraction is accumulated in
the created dense and highly excited matter, that implies a
strong rotational motion of this matter. This matter is conventionally 
associated with a (participant) fluid because it is successfully described 
by the fluid dynamics. 
Such rotation leads to a strong vortical structure inside the produced fluid. 
Local fluid vorticity induces a preferential orientation of spins of emitted
particles through spin-orbit coupling. The STAR Collaboration at 
the Relativistic Heavy-Ion Collider 
discovered the global polarization of emitted $\Lambda$ hyperons, which
indicated fluid vorticity of $\omega \approx (9\pm 1) 10^{21}$ s$^{-1}$ \cite{STAR:2017ckg}. 
This result exceeds the vorticity of all ever known fluids
in nature. This discovery have opened
an entirely new direction of research in heavy-ion physics.

Until recently, two main alternative mechanisms of this polarization were considered.  
The minor part of applications to heavy-ion collisions was based on
the chiral vortical effect  
\cite{Vilenkin:1980zv,Son:2004tq,Gao:2012ix,Sorin:2016smp,Baznat:2017jfj,Sun:2017xhx,Ivanov:2020qqe},  
while the major part was performed within   
thermodynamic approach in terms of hadronic degrees of freedom  
\cite{Becattini:2013fla,Becattini:2016gvu,Fang:2016vpj}.
In the present paper, we discuss this thermodynamic approach. 
The key quantity of the thermodynamic approach is thermal vorticity 
   \begin{eqnarray}
   \label{therm.vort.}
   \varpi_{\mu\nu} = \frac{1}{2}
   (\partial_{\nu} \beta_{\mu} - \partial_{\mu} \beta_{\nu}), 
   \end{eqnarray}
where $\beta_{\mu}=u_{\mu}/T$, 
$u_{\mu}$ is collective local four-velocity of the matter,  and $T$ is local temperature.  
The corresponding mean spin vector of $\Lambda$  particles with four-momentum $p$, 
produced around point $x$ on freeze-out hypersurface is
   \begin{eqnarray}
\label{xp-pol}
 S_{\varpi}^\mu(x,p)
 =-\frac{1}{8m} [1-f(x,p)] \: \epsilon^{\mu\nu\alpha\beta} p_\nu,  
  \varpi_{\alpha\beta}(x) 
   \end{eqnarray}
where $f(x,p) = 1/\{\exp[(u_{\nu}p^{\nu}-\mu)/T]+1\}$ 
is the Fermi-Dirac distribution function, $m$ is mass of the $\Lambda$ hyperon and 
$\mu$ is the baryon chemical potential. 
Numerous simulations of the polarization in heavy-ion collisions were based on 
this expression for the mean spin vector, see recent reviews 
\cite{Becattini:2020ngo,Karpenko:2021wdm,Huang:2020dtn}.

However, there are other contributions to the mean spin vector. 
A meson-field induced contribution to the global polarization  
was proposed in Ref. \cite{Csernai:2018yok} primarily to explain the observed 
$\Lambda$-$\bar{\Lambda}$ splitting in the global polarization. 
This meson-field term ($S_V^\mu$) gives additional to $S_{\varpi}^\mu$ contribution. 
It was implemented in Refs.  \cite{Xie:2019wxz,Ivanov:2022ble}. 

The thermal-vortical and meson-field terms are the only contributions to the particle polarization, 
if the thermal equilibrium is {\it global}. However, the fluid dynamics used to model 
the heavy-ion collisions implies the {\it local} thermal equilibrium.

It was recently realized \cite{Becattini:2021suc,Liu:2021uhn,Liu:2020dxg}
that there are other additional contributions to the mean spin vector, if 
the thermal equilibrium is {\it local}. These are 
the so-called thermal-shear ($S_{\xi}^\mu$) and spin-Hall ($S_{\zeta}^\mu$) contributions: 
\begin{eqnarray}
   \label{S-shear}
S_{\xi}^\mu (x,p) 
& = &  \frac{1}{4m} [1-f(x,p)]
\epsilon^{\mu\nu\alpha\beta}\frac{p_{\nu} n_{\beta}p^{\rho}}{(n\cdot p)} \xi_{\rho\alpha}, 
\\
   \label{S-Hall}
S_{\zeta}^\mu (x,p)
& = &  \frac{1}{4m} [1-f(x,p)]
\epsilon^{\mu\nu\alpha\beta}\frac{p_{\alpha}n_{\beta}}{(n\cdot p)}\partial_{\nu} \zeta,
\end{eqnarray}
where  $\zeta=\mu/T$, 
\begin{eqnarray}
   \label{shear}
  \xi_{\mu\nu} &=& \frac{1}{2} \left( \partial_\mu \beta_\nu + \partial_\nu \beta_\mu \right),  
\end{eqnarray}
is the thermal-shear tensor, 
and $n$ is a four-vector that is the main subject of the discussion below. 
In Ref. \cite{Becattini:2021suc}, the $n$ four-vector is defined as 
the time direction in the center-of-mass frame of colliding nuclei: 
$n_{\beta}=\hat{t}_{\beta}=(1,0,0,0)$. Only the shear term was considered in 
Ref. \cite{Becattini:2021suc}. Simulations with this choice of $n$ 
were performed in Ref. \cite{Becattini:2021iol}. In Refs. \cite{Liu:2021uhn,Liu:2020dxg}, 
the $n$ four-vector is identified with the four-velocity: $n_{\beta}=u_{\beta}$.
This choice was implemented in Refs. 
\cite{Fu:2021pok,Ryu:2021lnx,Yi:2021ryh,Sun:2021nsg,Fu:2022myl,Alzhrani:2022dpi,Wu:2022mkr}%
\footnote{ 
Definitions of the (thermal-)shear contribution are somewhat different in different papers. 
For example, by definition of the shear contribution in Ref. \cite{Wu:2022mkr}, the inverse temperature
in the thermal-shear tensor (\ref{shear}) is taken out from under the derivative sign
and also the term $(1/T)u_\mu(u\cdot\partial) u_\nu$ is subtracted from $\xi_{\mu\nu}$.
The thermal-shear contribution takes the form of Eq. (\ref{S-shear}), if 
the shear and ``acceleration'' contributions in terms of Ref. \cite{Wu:2022mkr} are combined.   
}. 
In the mid-rapidity region these choices are very close because $u_{\beta}\approx \hat{t}_{\beta}$. 
However, at forward-backward rapidities, which are relevant 
to fixed-target polarization measurements 
\cite{STAR:2021beb,Okubo:2021dbt,HADES:SQM2021}, 
they may significantly differ.

These different choices of $n_{\beta}$ are related to different approximations made 
within different approaches:  
the field-theoretical approach \cite{Becattini:2021suc} and 
the quantum kinetic approach \cite{Liu:2021uhn,Liu:2020dxg}. 
In Ref. \cite{Becattini:2021suc}, 
the $n_{\beta}$ vector is taken as 
the best approximation to the unit vector perpendicular to the freeze-out hypesurface, 
where the particle polarization is evaluated.  $n_{\beta}=\hat{t}_{\beta}$ is indeed a 
reasonable choice at the midrapidity. In Refs. \cite{Liu:2021uhn,Liu:2020dxg}, 
all the derivations are performed in the rest frame of the fluid element, therefore 
the normal vector to the corresponding hypesurface is $n_{\beta}=u_{\beta}$.

In this paper, we consider consequences of these different choices at the example of the 
global polarization of $\Lambda$-hyperons. 
The global polarization is chosen because it allows a significant advance in the analytical treatment, 
if momentum acceptance is disregarded, and because it is relevant to fixed-target polarization measurements 
at moderately relativistic energies \cite{STAR:2021beb,Okubo:2021dbt,HADES:SQM2021},   
if treated accordingly to Ref. \cite{Ivanov:2022ble}.

The polarization of the $\Lambda$ hyperon is measured in
its rest frame, therefore the $\Lambda$ polarization is 
   \begin{eqnarray}
   \label{P_L-rest}
  P^\mu(x,p) =  S^{*\mu}(x,p)/S_{\Lambda},  
   \end{eqnarray}
where $S_{\Lambda}=$ 1/2 is the spin of the $\Lambda$ hyperon,  
 $S^\mu=S_\varpi^\mu+S_V^\mu+S_{\xi}^\mu+S_{\zeta}^\mu$, and
$S^{*\mu}$ is  the mean $\Lambda$-spin vector its rest frame 
   \begin{eqnarray}
\label{S-rest}
 {\bf S}^*(x,p)
 = {\bf S} - 
 \frac{\hat{t} \cdot S}{\hat{t}\cdot p+m}
 {\bf p}_{\Lambda} 
 \stackrel{def}{=}
{\bf S} - \Delta{\bf S}, 
   \end{eqnarray}
where we used relation ${\bf p} \cdot {\bf S}=p^0 S^0$ and 
the above $\hat{t}$ definition, $S^0=\hat{t}\cdot S$ , to modify the conventional expression for $S^{*\mu}$. 
The zeroth component of $S^{*\mu}$  identically vanishes.

The global polarization implies averaging $P^\mu(x,p)$ over the freeze-out surface ($x$) 
and momenta ($p$), the latter within certain experimental acceptance. For a moment, we release 
the acceptance constraints. Then the global polarization reads 
   \begin{eqnarray}
\label{polint}
\langle  P^\mu\rangle
 = \frac{\int (d^3 p/p^0) \int_\Sigma d \Sigma \: s_\sigma p^\sigma
f(x,p)   P^\mu (x,p)}
 {\int (d^3 p/p^0) \int_\Sigma d\Sigma \: s_\sigma p^\sigma \, f(x,p)} . 
   \end{eqnarray}
where $d\Sigma_\sigma =d\Sigma \: s_\sigma$ is the aria of the surface element multiplied by 
normal vector to this element, $s_\sigma$. 
In fact, we are interested in $y$ component of $P^\mu$
because the global polarization is directed orthogonally to the reaction plane ($xz$).

Let us first perform integration over $p$. Then the denominator reads
   \begin{eqnarray}
\label{polint-denominator}
 \int \frac{d^3 p}{p^0} \int_\Sigma d\Sigma_\sigma p^\sigma \, f(x,p) 
=  \int_\Sigma d\Sigma_\sigma u^\sigma \rho_\Lambda
   \end{eqnarray}
where $\rho_\Lambda$ is the proper density of $\Lambda$'s. 
The $p$ integration of the thermal-vorticity, shear  and spin-Hall contributions in the 
numerator results in 
   \begin{eqnarray}
\label{polint-vort}
&&\int \frac{d^3 p}{p^0} s_\sigma p^\sigma f S_{\varpi}^\mu   
\cr
&\propto&
\epsilon^{\mu\nu\alpha\beta} \varpi_{\alpha\beta}
[A_{\varpi}  s_\sigma u^{\sigma} u_{\nu}  
+ D_{\varpi} s_{\nu}], 
\\
\label{polint-d-vort}
&&\int \frac{d^3 p}{p^0} s_\sigma p^\sigma f \Delta S_{\varpi}^\mu   
\cr
&\propto&
\epsilon^{0\nu\alpha\beta} s_\sigma  \varpi_{\alpha\beta} 
[\Delta A_{\varpi}^{\mu\sigma}  u_{\nu}  + 
\Delta D_{\varpi}^{\mu}  g^\sigma_{\nu}  + \Delta D_{\varpi}^{\sigma} g^\mu_{\nu} ],
\hspace*{6mm}
\\
\label{polint-shear}
&&\int \frac{d^3 p}{p^0} s_\sigma p^\sigma f S_{\xi}^\mu   
\cr
&\propto&
\epsilon^{\mu\nu\alpha\beta} n_{\beta} \xi_{\rho\alpha}
[A_{\xi}^{\sigma\rho}  s_\sigma u_{\nu}  
+ D_{\xi}^{\rho} s_{\nu}], 
\\
\label{polint-d-shear}
&&\int \frac{d^3 p}{p^0} s_\sigma p^\sigma f \Delta S_{\xi}^\mu   
\cr
&\propto&
\epsilon^{0\nu\alpha\beta}s_\sigma n_{\beta} \xi_{\rho\alpha}
[\Delta A_{\xi}^{\mu\sigma\rho}  u_{\nu} + \Delta D_{\xi}^{\mu\rho} g^{\sigma}_{\nu}+ \Delta D_{\xi}^{\sigma\rho} g^{\mu}_{\nu}], 
\hspace*{5mm}
\\
\label{polint-Hall}
&&\int \frac{d^3 p}{p^0} s_\sigma p^\sigma f S_{\zeta}^\mu 
\propto
\epsilon^{\mu\nu\alpha\beta}s_\sigma A_{\zeta}^{\sigma}  u_{\alpha}  n_{\beta}\partial_{\nu}\frac{\mu}{T},
\\
\label{polint-d-Hall}
&&\int \frac{d^3 p}{p^0} s_\sigma p^\sigma f \Delta S_{\zeta}^\mu 
\cr
&\propto&
\epsilon^{0\nu\alpha\beta}s_\sigma[\Delta A_{\zeta}^{\mu\sigma}u_{\alpha} +
\Delta D_{\zeta}^{\mu}  g^{\sigma}_{\alpha} + \Delta D_{\zeta}^{\sigma} g^\mu_{\alpha}]  
n_{\beta}\partial_{\nu}\frac{\mu}{T}.
   \end{eqnarray}
The tensor (vector) structure of $A$, $\Delta A$, $D$  and $\Delta D$ functions is built from available four vectors, 
$u$ and $n$, and also $\hat{t}$
for $\Delta A$ and $\Delta D$. This is a general structure of different contributions to the numerator
of Eq. (\ref{polint}) without specifying dependence on $T$ and $\mu$. Nevertheless, 
Eqs. (\ref{polint-vort})-(\ref{polint-d-Hall}) give us possibility to draw certain conclusions on the global polarization:

\begin{itemize}
	\item 
If $n^\mu=u^\mu$ \cite{Liu:2021uhn,Liu:2020dxg}, the spin-Hall contribution to the global polarization 
due to $S_{\zeta}^\mu$ is identically zero, 
as a result of convolution $\epsilon^{\mu\nu\alpha\beta} u_{\alpha}  n_{\beta}$.  

Nevertheless, a nonzero spin-Hall contribution to the global polarization was reported in Refs. \cite{Ryu:2021lnx,Alzhrani:2022dpi}. 
This was a result of the boost term, $\Delta S_{\zeta}^\mu$, and imposed momentum acceptance.  
Note that the boost $\Delta S$ correction is usually small compared to the main contribution 
\cite{Xie:2019wxz}, of course, if the main contribution is not identically zero. 

	\item 
If $n^\mu=u^\mu$ \cite{Liu:2021uhn,Liu:2020dxg} and also $s^\mu=u^\mu$, only the term resulting 
from the boost of the 
mean spin vector to the $\Lambda$ rest frame, $\Delta{\bf S}$, see (\ref{S-rest}), 
gives nonzero shear contribution to the global polarization. Moreover, this is only one of the three terms 
in Eq. (\ref{polint-d-shear}). 

At the same time, if $n^\mu=\hat{t}^\mu$ \cite{Becattini:2021suc}, ${\bf S}$ is nonzero, while the contribution 
from the boost term ($\Delta{\bf S}$) vanishes.  
Therefore, the global shear polarization is expected to be very different in $n^\mu=u^\mu$ and 
$n^\mu=\hat{t}^\mu$ cases. 

	\item 
The global thermal-vorticity polarization, Eqs. (\ref{polint-vort}) and (\ref{polint-d-vort}),
only moderately depends on the type, i.e. $s^\mu$, of the freeze-out surface.

\end{itemize}

In conclusion, alternative definitions of the thermal-shear contribution to the 
polarization in heavy-ion collisions, \cite{Becattini:2021suc} on the one hand and \cite{Liu:2021uhn,Liu:2020dxg}
on the other, 
result in very different corrections to  
the global $\Lambda$ polarization averaged over wide range of momenta. 
The spin-Hall contribution to the polarization, defined accordingly to 
Refs. \cite{Liu:2021uhn,Liu:2020dxg}, results in identicaly zero correction to the global $\Lambda$ polarization, if averaged 
over all momenta of $\Lambda$'s. Only application of restrictive momentum acceptance 
and the boost (to $\Lambda$ rest frame) correction 
results in nonzero 
global spin-Hall polarization. If the spin-Hall contribution were defined similarly 
to Ref. \cite{Becattini:2021suc}, 
the global spin-Hall polarization would be non-zero even without any acceptance and the boost correction.

This work was supported by MEPhI within the Federal Program "Priority-2030".

\end{document}